\documentclass[twoside,reqno]{HERON}
\usepackage{colordvi}
\usepackage{graphicx}
\usepackage{url}
\usepackage{amssymb,amsmath,amscd}
\usepackage{times}
\usepackage{makeidx}
\usepackage{here}
\begin{document}

\title{{$ \gamma- $rigid triaxial nuclei in the presence of a minimal length via a quantum perturbation method}}
\runningheads{Preparation of Papers for Heron Press Science Series
Books}{S. Ait Elkorchi, M. Chabab, A. El Batoul, A. Lahbas, M. Oulne}
%\begin{start}
%\author{S.Ait Elkorchi.}, \coauthor{ M. Chabab}, \coauthor{ A. El Batoul}, %\coauthor{  M. Oulne}
%\index{S.Ait Elkorchi.}
%\index{ M. Chabab.}
%\index{ A. El Batoul.}
%\address{Academic Affiliation, Zip Code City, State}{1}
%\address{Academic Affiliation, Zip Code City, State}{2}
\begin{start}

	\author{S. Ait Elkorchi}{1}, \coauthor{ M. Chabab}{1}, \coauthor{ A. El Batoul}{1},\coauthor{ A. Lahbas}{1}, \coauthor{  M. Oulne}{1}
	
	\index{Author, F.A.}

	\address{ High Energy Physics and Astrophysics Laboratory, Faculty of Sciences Semlalia, Cadi Ayyad University, P. O. B. 2390, Marrakech 40000, Morocco}{1}

\begin{Abstract}
	In this work, we derive a closed solution of the Shr$ \ddot{o} $dinger equation for Bohr Hamiltonien within the minimal length formalism. This formalism is inspired by Heisenberg algebra and a generlized uncertainty principle (GUP), applied to the geometrical collective Bohr- Mottelson model (BMM) of nuclei by means of deformed canonical commutation relation and the Pauli-Podolsky prescription. The problem is solved by means conjointly of asymptotic iteration method (AIM) and a quantum perturbation method (QPM) for transitional nuclei near the critical point symmetry Z(4) corresponding to phase transition  from prolate to $ \gamma-rigid $ triaxial shape. A scaled Davidson potentiel is used as a restoring potential in order to get physical minimum.
	The agreement between the obtained theoretical results and the experimental data is very satisfactory.
	
\end{Abstract}
\end{start}
\section{Introduction}
Critical point symmetries in nuclear structure are recently receiving considerable 
attention\cite{a,c} since they provide parameter-free solutions. The pioneering ones amid them were E(5) \cite{a, bohr}, X(5) \cite {b}
and Z(5)\cite{Z(5)} corresponding to shape phase transitions from U(5) to O(6),
U(5) to SU(3) and from axial to triaxial shapes respectively, with the recent addition of Y(5) \cite{c} related to the transition from prolate to oblate shape
. Later, a $ \gamma $-rigid (with $ \gamma = 0 $)
version of X(5), called X(3), has been introduced \cite{X(3)}. In
the same way, other CPS have been developped like for example Z(4) (with 
$ \gamma =\pi/6 $) the gamma rigid version of Z(5)
corresponding to shape phase transitions from prolate to
triaxial symmetry \cite{8, ZZZ, Z(4),Z(44)}. From a structural point of view, the collective Bohr Mottelson 
 represents a sound frame work to describe many properties of the quadrupole collective dynamics in nuclei \cite{bohr}. Its formulation has the ability to describe both rotational and vibrational modes. On the other hand, recently, a great interest has been consecrated to the quantum mechanical problems related to a generalized modifed commutation relations involving a minimal
 length or generalized uncertainty principle \cite{uncer, unc}.\\
  In the present work we focuse on the study of the quadrupole
  collective states in $ \gamma $-rigid case, by modifying Davydov-Chaban Hamiltonian in the framework of the minimal length formalism \cite{ML} with Davidson potential \cite{Davidson} for$ \beta $-vibrations. The model is conventionally called Z(4)-D-ML. The organization of this paper is as follows: in
 section 2, we present the Z(4)-D-ML model with the quantum perturbation method which requires the study of the model in the absence and in the presence of the minimal length, presented in sections 3 and 4 respectively. Finally, section 5 is devoted to the numerical results and brief discussion for energy spectrum of some triaxial-rigid nuclei, while section 6 contains our conclusions.
  \section{ Z(4)-D-ML with the quantum perturbation method}
 
  In the Z(4) model the $\gamma $ variable is “frozen” to $\gamma= \pi/6 $ and only four variables are involved; three Euler angles $ (\theta_{1}, \theta_{2}, \theta_{3}) $, which obviously define the orientation of the intrinsic principal axes in the laboratory frame and the deformation parameter.
  So, in the frame of the Bohr-Mottelson model \cite{bohr,bohr1}, the corresponding eigenvalue problem reduced to that of the Davydov-Chaban hamiltonian \cite{8}. Therefore, we aimed to study the minimal length effect on energy spectrum in the context of $ \gamma $-rigid nuclei Z[4].
  \begin{equation}
  H_{DC}=	-\dfrac{\hbar^{2}}{2B_{m}}\left[ \dfrac{1}{\beta^{3}}\dfrac{\partial}{\partial\beta}\beta^{3}\dfrac{\partial}{\partial\beta}-\dfrac{1}{4\beta}\sum^{3}_{k=1}\dfrac{Q_{k}^{2}}{sin^{2}(\gamma-\frac{2\pi}{3}k)}\right]+U(\beta),
  \label{1}
  \end{equation}
  Where $\beta $ and $  \gamma $ are the usual collective coordinates \cite{bohr}, $  Q_{k} (k = 1, 2, 3) $ are the components of angular
  momentum and $ B_{m} $ is the mass parameter. In this
  Hamiltonian $\gamma $ is treated as a parameter and not as a
  variable.\\
   By employing the mathematical formulation, including the minimal
 length concept, presented in the original paper\cite{ML}, the
 collective equation of eigenstates, up to the first order of $ \alpha $, is written as follows:
 \begin{equation}
\left(  -\dfrac{\hbar^{2}}{2B_{m}}\Delta+\dfrac{\alpha \hbar^{4}}{B_{m}}\Delta^{2}+V(\beta)-E_{n,L}\right) \psi(\beta, \theta_{i}) =0,
\label{ml}
 \end{equation}
 where \\
  \begin{equation}
  \Delta=\dfrac{1}{\beta^{3}}\dfrac{\partial}{\partial\beta}\beta^{3}\dfrac{\partial}{\partial\beta}-\dfrac{\Delta_{\theta}}{4\beta} 
  \end{equation}
   \begin{equation}
   \Delta_{\theta}=\sum^{3}_{k=1}\dfrac{Q_{k}^{2}}{sin^{2}(\gamma-\frac{2\pi}{3}k)} 
   \end{equation}
 and  $\theta _{i} (i= 1, 2, 3) $ are the Euler angles.
 This equation can be simplifed by introducing an auxiliary
 wave function\cite{ML}:
 \begin{equation}
 \psi(\beta, \theta_{i})=(1+2\alpha\hbar^{2}\Delta)\xi(\beta, \theta_{i}).
 \label{auxi}
 \end{equation}
 Thus, we obtain the following differential equation satisfied by $ \xi(\beta, \theta_{i}) $
 \begin{equation}
 \left[ (1+4B_{m}\alpha(E-V(\beta)))\Delta+\dfrac{2B_{m}}{\hbar^{2}}(E_{n,L}-V(\beta))\right]\xi(\beta, \theta_{i}) =0.
 \end{equation}
 This equation can be simplified by using the usual following factorization
 \begin{equation}
 \xi(\beta, \theta_{i})=\Phi(\beta)\chi(\theta_{i} ).
 \end{equation}

 	The separation of variables leads to two equations : one 	depending only on the $ \beta $ variable and the other depending on the $ \gamma $ and the Euler angles :
 	\begin{equation}
 	\left[ \dfrac{1}{\beta^{3}}\dfrac{\partial}{\partial\beta}\beta^{3}\dfrac{\partial}{\partial\beta}-\dfrac{\Lambda}{\beta^{2}}+\dfrac{2B_{m}}{\hbar^{2}} \bar{k}(E_{n,L},\beta)\right] \Phi(\beta)=0,
 	\label{eq3}
 	\end{equation}
 	where \\
 	\begin{equation}
 	\bar{K}(E_{n,L},\beta) =\left( \dfrac{E_{n}-V(\beta)}{1+4B_{m}\alpha(E_{n,L}-V(\beta)}\right)
 	\label{K}
 	\end{equation}
 	 $ n$ is the radial quantum number.
 	\begin{equation}
 	[\dfrac{1}{
 		4}\sum^{3}_{k=1}\dfrac{Q_{k}^{2}}{sin^{2}(\gamma-\frac{2\pi}{3}k)}-\Lambda]\chi(\theta_{i} )=0.
 	\end{equation}
 	In the case of $ \gamma = \pi/6 $, the last equation takes the form \cite{10}:
 	\begin{equation}
 	[\dfrac{1}{4}(Q_{1}^{2}+4Q_{2}^{2}+4Q_{3}^{2})-\Lambda]\chi(\theta_{i} )=0.
 	\label{4}
 	\end{equation}
 		This equation has been solved by Meyer-ter-Vehn
 		\cite{10}, the eigen functions being:
 		 	\begin{equation} \chi(\theta_{i} )= \chi(\theta_{i} )_{\mu,\omega}^{L} 
 		  =\sqrt{\dfrac{2L+1}{16\pi^{2}(1+\delta_{\omega},0)}}\times [D^{L}_{\mu,\omega} (\theta_{i})+(-1)^{L}D^{L}_{\mu,-\omega} (\theta_{i})].
 		  \end{equation}
 		 Here, $ D^{L}_{\mu,\omega} (\theta_{i}) $ represents the Weigner function of the Euler angles, L are the eigenvalues of angular momentum,
 		 while $  \omega $ and $ \mu $ are the eigenvalues of the projections of angular momentum on the body-fixed ˆx-axis
 		 and the laboratory fixed ˆz-axis, respectively.\cite{10} with\\
 			\begin{equation}
 			\Lambda=\Lambda_{L,\omega}=L(L+1)-\frac{3}{4}\omega^{2}.
 			\end{equation}
 			
 	 Thanks to the smallness of the parameter $ \alpha $, by expanding \eqref{K} in power series of $ \alpha $, one can obtain diferent order approximations of the standard model Z(4)-ML. At the first order approximation, as it has been done recently in \cite{qpm} \eqref{K} becomes:\\
 	 
 	 $  
 	 \bar{K}(E_{n,L}-V(\beta))\approx(E_{n,L}-V(\beta))(1-4B_{m}\alpha(E_{n,L}-V(\beta)))
 	 $	
 	 \begin{equation}
 	 \quad=E_{n,L}-V(\beta)-4B_{m}\alpha(E_{n,L}-V(\beta))^{2}.
 	 \end{equation}	
 	 In what concerns the $ \beta $ degree of freedom, we will consider the Davidson potential chosen to be of the following form:
 	 \begin{equation}
 	 V(\beta)=a\beta^{2} +
 	 \dfrac{b}{\beta^{2}}, \quad \beta_{0}=(\dfrac{b}{a})^{\frac{1}{4}}
 	 \end{equation}
 	 where a and b are two free scaling parameters, and $\beta_{0} $ represents the position of the minimum of the potential. The special case of $b=0$ ($\beta_{0}=0$) corresponds to the simple harmonic oscillator.\\
 	 The diferential equation
 	 \eqref{eq3} was solved exactly, with an infinite square
 	 well like potential, within the standard model \cite{com} but 
 	  it is not soluble analytically for the Davidson-type potential. However, the quantum perturbation theory one of its familiar forms, dubbed the quantum perturbation method (QPM), is used to obtain approximate solutions for all values of angular momentum L \cite{qpm}.
 	  
\section{Z(4) model with Davidson potential Z(4)-D ($ \alpha=0 $)}
It is preferable to write the equation 
in a Schr$ \ddot{o} $dinger picture. This is realized by changing the
wave function as $ \Phi(\beta)= \beta^{-\frac{3}{2}}f(\beta) $. However one obtains an equation which resembles the radial Schr$ \ddot{o} $dinger
equation for an isotropic Harmonic Oscillator acting in
four-dimensional space:
\begin{equation}
\dfrac{d^{2}}{d\beta^{2}}f(\beta)+\left[ \dfrac{2B_{m}E_{n,L}}{\hbar^{2}}-\dfrac{2B_{m}a}{\hbar^{2}}\beta^{2}
-\dfrac{L(L+1)+\frac{3}{4}(1-\alpha^{2})+\dfrac{2bB_{m}}{\hbar^{2}}}{\beta^{2}} \right]f(\beta)=0.
\end{equation}
 We define:\\
 \begin{center}
 $ \epsilon=\dfrac{2B_{m}E_{n,L}}{\hbar^{2}} $,	$\quad \omega=\dfrac{2B_{m}a}{\hbar^{2}} ,$\\
 \end{center}
  and
 \begin{equation}
\vartheta_{l,b}(\vartheta_{l,b}+1)=L(L+1)+\frac{3}{4}(1-\alpha^{2})+\dfrac{2bB_{m}}{\hbar^{2}} .
\label{16}
 \end{equation} 

However one obtains an equation which resembles to the Gol’dman and Krivchenkov Hamiltonian \cite{sol}

\begin{equation}
\dfrac{d^{2}}{d\beta^{2}}f(\beta)+\left[ \epsilon-\omega\beta^{2}-\dfrac{\vartheta_{l,b}(\vartheta_{l,b}+1)}{\beta^{2}}\right] f(\beta)=0.
\end{equation}
To solve this diferential equation via the asymptotic iteration
method (AIM)\cite{sol}, we propose the following
ansatz\cite{sol}:\\
\begin{equation}
f(\beta)=\beta^{1+\vartheta_{l,b}}+e^{-\frac{\sqrt{\omega}}{2}\beta^{2}}\Theta(\beta).
\end{equation}
Thus we obtain,

	\begin{equation}
	\dfrac{d^{2}\Theta(\beta)}{d\beta^{2}}+(\frac{2p}{\beta}-4q\beta)\dfrac{d\Theta(\beta)}{d\beta}+[\epsilon-2q(1+2p)]\Theta(\beta)=0,
	\end{equation}
where $ \quad $
	$q=\dfrac{\sqrt{\omega}}{2} $ and $ p=1+\vartheta_{l,b}$.\\
	After calculating $ \lambda_{0} $ and $ s_{0} $, by means of the recurrence relations\cite{sol}, we get the generalized formula of the reduced energy from the roots of the quantization condition
	
	\begin{equation}
	\epsilon=q[2+4p+ 8n] ; n=1,2,....
	\end{equation}
	from which, we obtain the energy spectrum :
	\begin{equation}
	E_{n,L}=\dfrac{\hbar^{2}}{2B_{m}}\epsilon=\sqrt{\dfrac{\hbar^{2}}{2B_{m}}}a[3+4n_{\beta}+2\vartheta_{l,b}].
	\end{equation}
	 From equation \eqref{16}, we get $\vartheta_{l,b} $ as a function of the total angular momentum L and the parameter b :
	 \begin{equation}
	\vartheta_{l,b}=-\dfrac{1}{2}+\dfrac{1}{2}\sqrt{4L(L+1)-3\alpha^{2}+8b+4}.
	 \end{equation}
 The physical solutions to \eqref{eq3} are obtained as:
	
\begin{equation}
\Phi(\beta)=N_{n\beta,L}\beta^{-\frac{3}{2}+p}e^{-q\beta^{2}}L^{p-\frac{1}{2}}_{n\beta}(2q\beta^{2}),
\end{equation}
where $  L^{p-\frac{1}{2}}_{n\beta}(2q\beta^{2}) $ denotes the associated Laguerre polynomials
and $ N_{n\beta,L} $ is a normalization constant.	
\section{Z(4) model with Davidson potential via a minimal length (Z(4)-D-ML)}
Here, we treat the additional term$ \dfrac{\alpha \hbar^{4}}{B_{m}}\Delta^{2} $ shown
in equation \eqref{ml} as a perturbation and then estimate its effect on the energy spectrum up to the first order of the perturbation theory. Hence, the energy spectrum can be written as:
\begin{equation}
E_{n,L}=E_{n,L}^{0}+\Delta E_{n,L},
\label{ener}
\end{equation}
where $ E_{n,L}^{0} $ is the unperturbed energy spectrum, and $ \Delta E_{n,L} $ the correction induced by the minimal length, given by:
\begin{equation}
\Delta E_{n,L}=\dfrac{\alpha \hbar^{4}}{B_{m}}<\psi^{0}(\beta,\theta_{i})\mid\Delta^{2}\mid \psi^{0}(\beta,\theta_{i})>,
\end{equation}
where $ \psi^{0}(\beta,\theta_{i}) $ are the eigenfunctions, solutions to the ordinary Schr$\ddot{o} $dinger equation ($ \alpha=0 $) .\\
So, the energy spectrum  can be expressed as \cite{qpm},
\begin{equation}
\Delta E_{n,L}=4B_{m}\alpha\left[ (E_{n,L}^{0})^{2}-2E_{n,L}^{0}<\psi^{0}\mid V(\beta)\mid \psi^{0}>+<\psi^{0}\mid V(\beta)^{2}\mid \psi^{0}>\right] .
\end{equation}
After substituting the Davidson potential (16), one obtains
\begin{equation}
\Delta E_{n,L}=4B_{m}\alpha\left[ (E_{n,L}^{0})^{2}+2ab-2E_{n,L}^{0} (a\overline{\beta^{2}}+b\overline{\beta^{-2}})+(a^{2}\overline{\beta^{4}}+b^{2}\overline{\beta^{-4}})\right] ,
\end{equation}
Where $ \overline{\beta^{i}} $ (i=2,- 2,4,- 4) are expressed as follows\\
\begin{equation}
\left\lbrace
\begin{aligned}
	\overline{\beta^{2}}=\dfrac{4n+2\vartheta_{l,b}+3}{4q} \\
	 \overline{\beta^{-2}}=\dfrac{4q}{2\vartheta_{l,b}+1} \\
	 \overline{\beta^{4}}=\dfrac{4\vartheta_{l,b}^{2}+24n\vartheta_{l,b} + 24n^{2}+16\vartheta_{l,b}+36n+15}{16q^{2}} \\
\overline{\beta^{-4}}=\dfrac{16q^{2}(4n+2\vartheta_{l,b}+3)}{(2\vartheta_{l,b}+3)(4\vartheta_{l,b}^{2}-1)}
\end{aligned}
\right. 
\end{equation}
Details of $ \overline{\beta^{i}} $ calculations are given in \cite{qpm}.
\section{Numerical examination}
\begin{figure}[H]
 	%\centering
 	\centering
 	\includegraphics[scale=0.48]{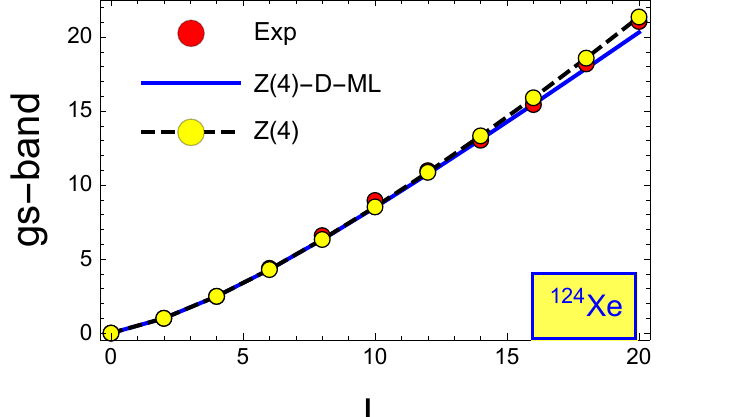}
 	\includegraphics[scale=0.48]{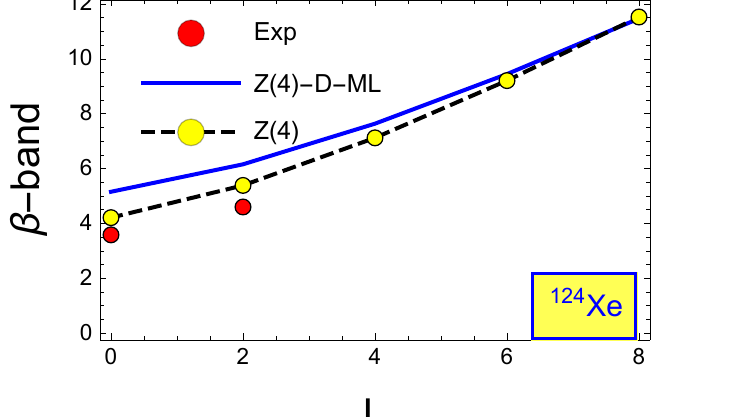}
 	\includegraphics[scale=0.48]{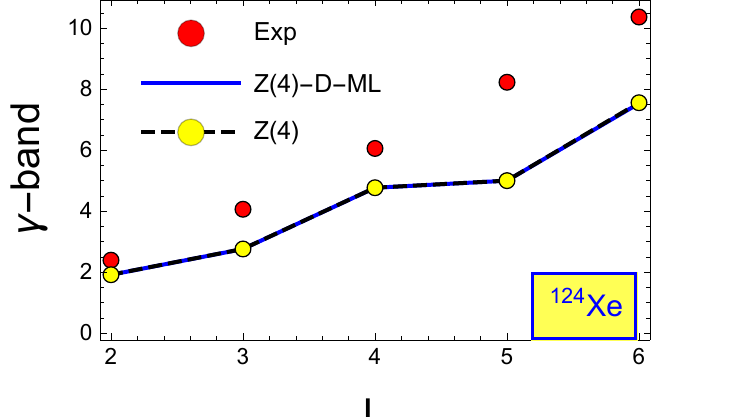}
 	\centering
 	\includegraphics[scale=0.48]{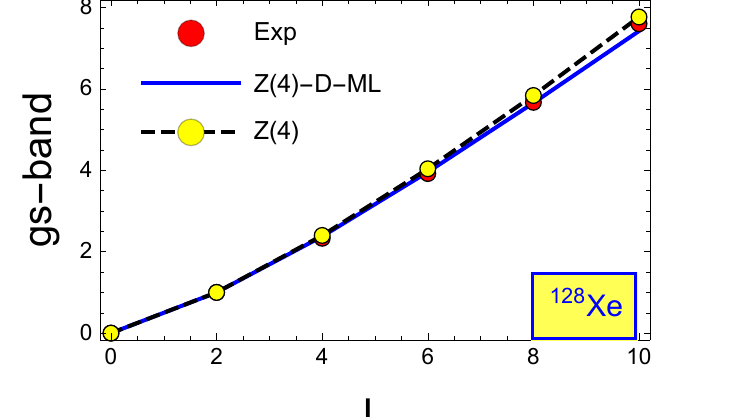}
 	\includegraphics[scale=0.48]{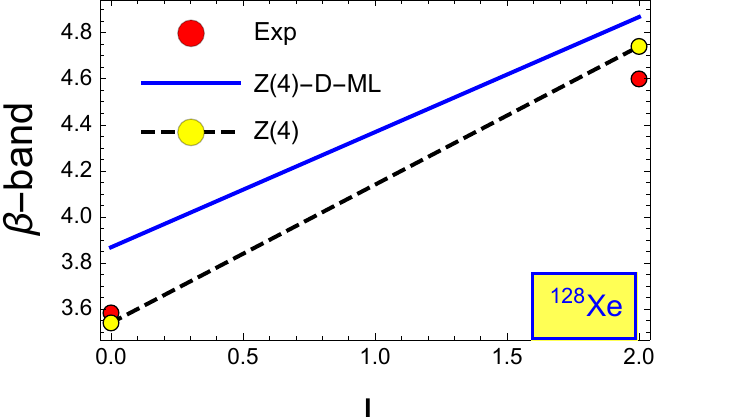}
 	\includegraphics[scale=0.48]{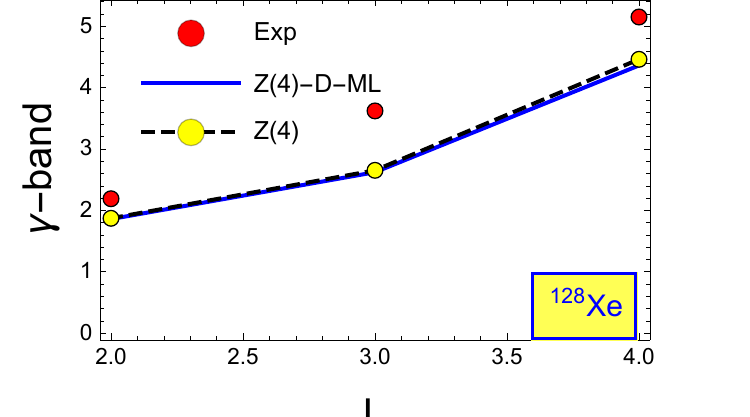}
 	
 	\centering
 	\includegraphics[scale=0.48]{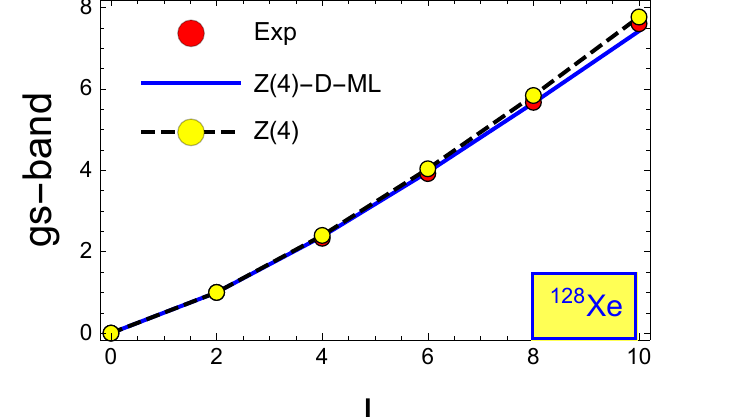}
 	\includegraphics[scale=0.48]{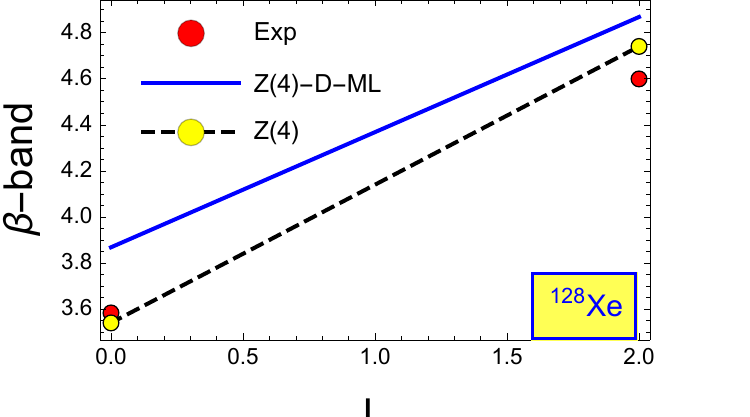}
 	\includegraphics[scale=0.48]{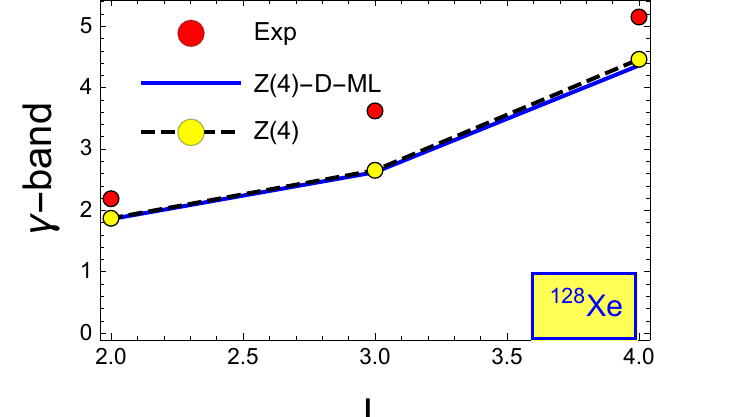}	
 	\centering
 	\includegraphics[scale=0.48]{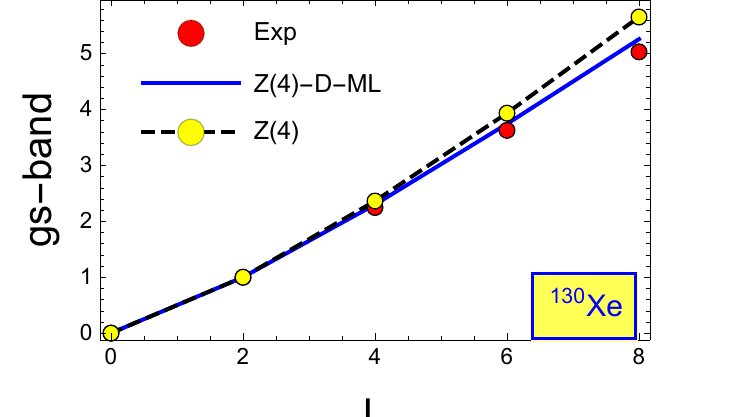}
 	\includegraphics[scale=0.48]{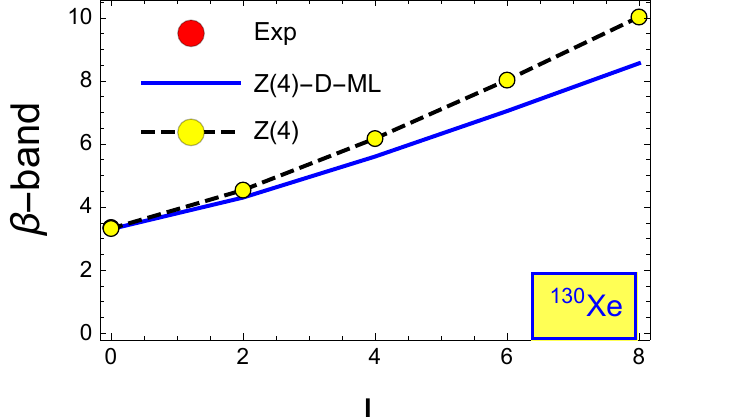}
 	\includegraphics[scale=0.48]{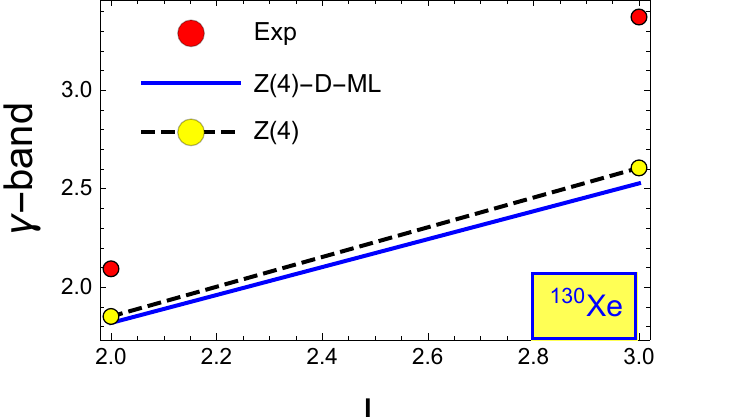}
 	\centering
 	\includegraphics[scale=0.48]{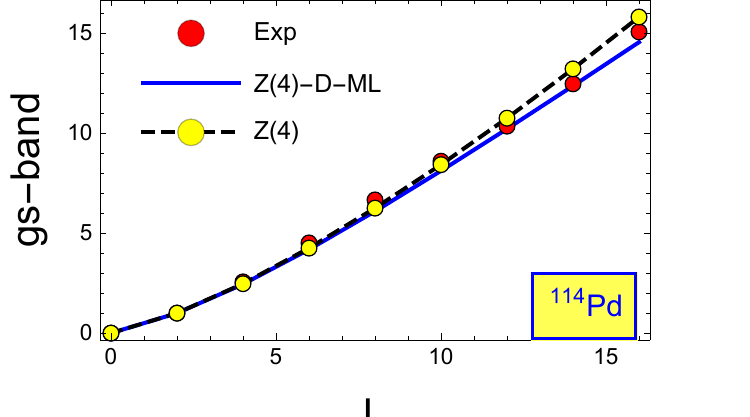}
 	\includegraphics[scale=0.48]{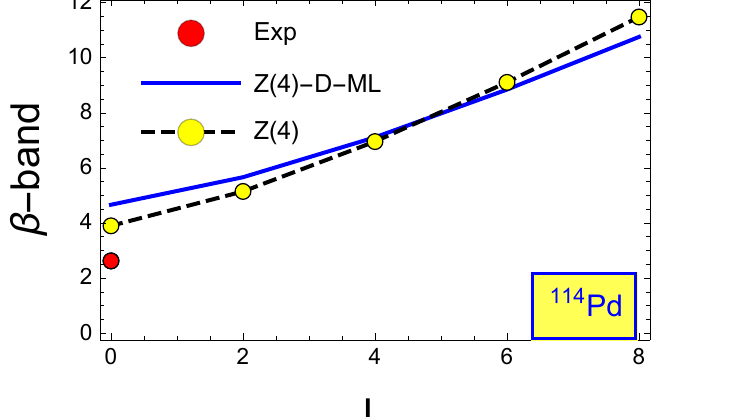}
 	\includegraphics[scale=0.48]{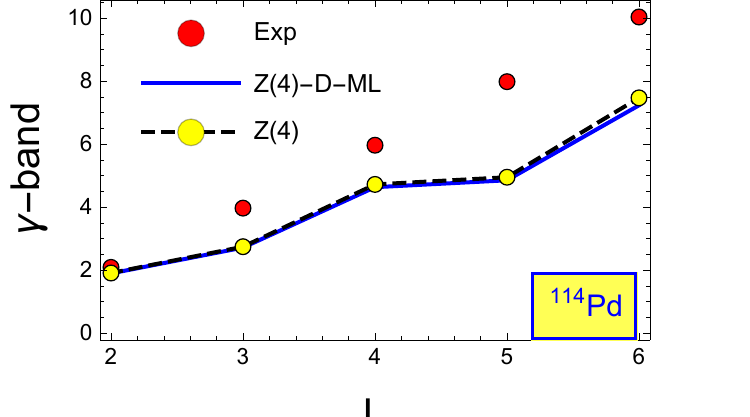}		
 	\caption{Comparison of the Z(4)-ML-D predictions for (normalized) energy levels to experimental
 		data for $ ^{124}Xe $ ,  $ ^{126}Xe $, $ ^{128}Xe $, $^{130}Xe  $, $  ^{114}Pd$}\label{f01}
 \end{figure}

The model established in this work, called Z(4)-D-ML, is adequate for description of
$ \gamma $-rigid nuclei for which the $  \gamma $ parameter is fixed to $ \gamma=\pi/6 $. Basically, the energy levels of the
ground state band as well as of the vibrational bands are characterized by the principal quantum number $ n_{\beta} $ and $ n_{\omega} $, respectively.
with $ n_{\omega} $ is the wobbling quantum number \cite{10,bohr2} $n_{\omega} = L- \omega $.
 We briefly recall a few interesting low-lying bands which are
 classified by the quantum numbers $ n_{\beta} $ and $ n_{\omega} $
 \begin{itemize}
 	 \item  The ground state band (gsb) with $ n_{\beta}=0 $ and $ n_{\omega}=0 $
 	 \item  The $\beta $ band with $ n_{\beta}=1 $ and $ n_{\omega}=0 $
 	 \item The$ \gamma $  
 	 band composed by the even L levels with $ n_{\beta}=0 $ and $ n_{\omega}=2 $
 	 and the odd L levels with  $ n_{\beta}=0 $ and $ n_{\omega}=1 $
 	 \item For our subsequent calculations, we define the energy ratios as:\\
 	 $ R(n_{\beta},L, n_{\omega}) =\dfrac{E_{n_{\beta},L, n_{\omega}}-E_{0,0, 0}}{E_{0,2, 0}-E_{0,0, 0}} $
 	 \item The mentioned results are thus found to
 	 have the smallest deviations from the experimental data \cite{data}, evaluated by the quality	measure\\
 	  $ \sigma=\sqrt{(\sum_{i=1}^{N}(E_{i}^{exp}-E_{i}^{th})^{2})/{N}} ,$\\
 	  
 	  where N is the maximum number of levels.
 	 We have treated 32 nuclei among which are depicted in Figure~\ref{f01} those with good results
 \end{itemize}

 \begin{figure}
 	
 \end{figure}
 	 
 The comparison between Z(4)-D-ML theoretical predictions
 and experimental data\cite{data} of selected candidates regarding energy levels is
 visualized schematically in Fig.1. The agreement with experiment is very good
 for the ground state band and $ \beta $
 band, despite the fact that there is not much experimental data especially for the $\gamma $ band of these studied nuclei. As a result, one concludes that the Z(4)- D-ML is more suitable for describing the structural properties of nuclei having a structure in vicinity of the Z(4) limit.
 \begin{table}[h]
 	
 	\smallskip
 	\small\noindent\tabcolsep=9pt
 	%\tiny 
 	\begin{tabular}{ c c c c ccc }
 		\hline 
 		
 		\hline
 		\\[-8pt]
 		%	&\multicolumn{3}{c}{} & &\multicolumn{2}{c}{}  \\
 		%	\cline{2-4}
 		%	\\[-8pt]
 		nuclei \qquad&$ {\sigma_{D}}$ & $ {\sigma_{D-ML}}$&$ \beta_{0} $\\
 		\hline
 		\\[-8pt]
 		{$^{124}Xe$} \quad&0.588&0.44&0.82  \\
 		{$^{126}Xe$} \quad&0.36&0.36&0.61  \\
 		{$^{128}Xe$} \quad&0.44& 0.40&0.69 \\
 		{$^{130}Xe$} \quad&0.35& 0.35&0.62 \\
 		{$^{114}Pd$} \quad&0.84& 0.63&0.78 \\
 		
 	\end{tabular}
 	\caption{Standard deviation between experimental and theoretical results}
 \end{table}
 	\section{Conclusion}
 	 The idea of Z(4)-ML is already used and presented with the square wel potential \cite{com}, but this time it was used with the Davidson potential. However, the Hamiltonian of the system
 	 is not soluble analytically for a potential
 	 other than the square well. In order to overcome such a difficulty, in the present work we used, a quantum perturbation method  (QPM), to obtain approximate solutions for all values of angular momentum L.
 	 Therefore, closed-form analytical formula for the energy of the ground and vibrational bands was derived for trixial $ \gamma $-rigid nuclei within Davidson potential.  Our results indicate a better agreement with the experimental values, and reproduced well the best Z(4) condidate nuclei already obtained in the Xe region around A = 130 including the new one $ ^{ 114}Pd $.
 	\section*{Acknowledgements}
 	S. Ait Elkorchi would like to thank the	organizing committee for the hospitality and the wonderful scientific meet. Also, she acknowledges
 	the financial support (Type A) of Cadi Ayyad University.

\end{document}